# Infrared frequency-tunable coherent thermal sources

Hao Wang,[*] Yue Yang,[*] and Liping Wang[†]

School for Engineering of Matter, Transport & Energy, Arizona State University, Tempe, Arizona, 85287, USA

[*] Equal contribution.

[†] Corresponding author. Email: liping.wang@asu.edu, Phone: 1-404-727-8615

**Abstract**

In this work, we numerically demonstrate an infrared frequency-tunable selective thermal emitter made of graphene-covered silicon carbide (SiC) gratings. Rigorous coupled-wave analysis shows temporally-coherent emission peaks associated with magnetic polariton, whose resonance frequency can be dynamically tuned within the phonon absorption band of SiC by varying graphene chemical potential. An analytical inductor-capacitor circuit model is introduced to quantitatively predict the resonance frequency and further elucidate the mechanism for the tunable emission peak. The effects of grating geometric parameters, such as grating height, groove width and grating period, on the selective emission peak are explored. The direction-independent behavior of magnetic polariton and associated coherent emission are also demonstrated. Moreover, by depositing 4 layers of graphene sheets onto the SiC gratings, a large tunability of 8.5% in peak frequency can be obtained to yield the coherent emission covering a broad frequency range from 820 cm$^{-1}$ to 890 cm$^{-1}$. The novel tunable metamaterial could pave the way to a new class of tunable thermal sources in the infrared region.

## 1. Introduction

Infrared (IR) spectroscopy plays an important role in material analysis to provide information on chemical composition and bonds. IR sources with tunable frequencies are highly desired for IR spectroscopy, while tunable IR sources like Quantum Cascade Lasers [1-3] are expensive. As a pioneering work in coherent thermal sources, Greffet et al. demonstrated both temporal and spatial coherences of thermal emission from SiC gratings by exciting surface phonon polaritons (SPhP) [4]. Similarly, surface plasmon polaritons can be employed in metallic micro/nanostructures such as one-dimensional (1D) complex grating [5], 2D tungsten grating [6], and photonic crystals [7-9] for tailoring thermal emission. The cavity resonant mode excited in the so-called Salisbury screen [10] and Fabry-Perot cavity [11, 12] is another way to achieve coherent emission.

Recently, metamaterials [13] with exotic optical and radiative properties that cannot be obtained in naturally-occurring materials have also been proposed for IR emitters. Liu et al. demonstrated single and dual-band IR emitters with metallic cross-bar resonators [14]. Wang and Zhang reported the direct measurement of infrared coherent emission due to magnetic polaritons (MP) in film-coupled grating metamaterial microstructures under elevated temperatures up to 750 K [15]. Besides incandescent infrared sources [16], coherent emission has numerous promising applications in energy harvesting [17-19], chemical sensing [20], and radiative cooling [21]. Thermal emitters with tunable emitting frequencies in a broad spectral range are highly desired for IR spectroscopy and energy applications.

By employing tunable or phase transition materials, optical and radiative properties of metamaterials can be actively controlled with chemical, thermal and electrical approaches. By use of InSb whose carrier density can be adjusted by utilizing optical pump or changing surrounding temperature, the resonant frequency of split rings which was used as a tunable selective absorber can be varied [22]. Yttrium Hydride nanoantennas were proposed as switchable devices, in which the plasmonic resonance can be turned on and off upon phase change of $YH_x$ induced by hydrogen exposure [23]. The phase transition material vanadium oxide ($VO_2$) has been employed in thermally induced switchable [24] and tunable [25-27] metamaterial structures. Tunable mid-infrared metamaterial using Y-shape plasmonic antenna array on a $VO_2$ film was also demonstrated upon temperature variation [28]. Besides, Ben-Abdallah et al. [29] proposed a phase-change thermal antenna made of patterned $VO_2$ gratings that exhibits switchable thermal emission. Liquid crystals were also introduced for electrically tunable metamaterials [30, 31].

Graphene has been recently employed in the novel designs of switchable and tunable metamaterials, as its optical properties [32-34] vary with the chemical potential which can be tuned by chemical doping, voltage bias, external magnetic field, or optical excitation [35-37]. Tunable selective transmission has been investigated in patterned graphene ribbon arrays by actively exciting plasmonic resonances [38, 39]. One step further, Chu et al. introduced an active plasmonic switch with dynamically controlled transmission in both single and multi-layer graphene ribbon arrays [40]. In addition, tunable perfect absorbers were investigated with graphene ribbon array on dielectric spacer and metallic substrate [41, 42]. Fang et al. demonstrated tunable selective absorption in graphene disk arrays [43]. Enhanced light absorption was also observed in graphene layer integrated with a metamaterial perfect absorber [44]. However, graphene-based tunable coherent thermal emission has not been demonstrated yet.

In this work, we numerically design an infrared frequency-tunable thermal emitter whose spectrally-selective emission peak can be shifted by varying graphene chemical potential. Figure 1 schematizes the proposed tunable metamaterial structure, which is made of a graphene-covered 1D SiC grating array with period $\Lambda = 5$ μm, groove width $b = 0.5$ μm, and grating height $h = 1$ μm. SiC is chosen as the thermal emitter material due to its excellent high temperature stability. The SiC grating with submicron feature sizes considered here can be practically realized with advanced nanofabrication techniques such as electron-beam lithography or high-throughput low-cost nanoimprint, deep-UV, or laser interference lithography. The graphene layers can be deposited onto the grating layer from chemical vapor deposition. The tunable coherent emission in this study is achieved via the modulation of phonon-mediated MP condition by tuning the optical properties of graphene.

## 2. Theoretical Background

### 2.1. Optical Properties of Materials

Electrical permittivity of monolayer graphene at optical frequencies can be described by [32-34]

$$\varepsilon_G = \frac{i\sigma_G}{\omega\varepsilon_0 t_G} \quad (1)$$

where $\sigma_G$ is the conductivity of graphene, $t_G$ is the thickness of a single graphene layer, $\omega$ is the angular frequency, and $\varepsilon_0$ is the permittivity of vacuum. The conductivity of monolayer graphene is acquired by $\sigma_G = \sigma_D + \sigma_I$. Note that $\sigma_D$ corresponds to the intraband electron-photon scattering, while $\sigma_I$ is associated with interband electron transition. $\sigma_D$ and $\sigma_I$ can be calculated by:

$$\sigma_D = \frac{i}{\omega + i/\tau} \frac{2e^2 k_B T}{\pi \hbar^2} \ln\left(2\cosh\left(\frac{\mu}{2k_B T}\right)\right) \quad (2)$$

$$\sigma_I = \frac{e^2}{4\hbar}\left[G\left(\frac{\hbar\omega}{2}\right) + i\frac{4\hbar\omega}{\pi}\int_0^\infty \frac{G(\xi) - G(\hbar\omega/2)}{(\hbar\omega)^2 - 4\xi^2}d\xi\right] \quad (3)$$

where $G(\xi) = \sinh(\frac{\xi}{k_B T})/\left[\cosh\left(\frac{\mu}{k_B T}\right) + \cosh\left(\frac{\xi}{k_B T}\right)\right]$, $\tau$ is the relaxation time chosen as $10^{-13}$ s [45], $e$ is the elementary charge, $k_B$ is the Boltzmann's constant, h is the reduced Planck's constant, temperature $T$ is taken as 300 K, and $\mu$ is chemical potential of graphene.

On the other hand, dielectric functions of SiC are given by a Lorentz oscillator model as [46]:

$$\varepsilon_{SiC}(\nu) = \varepsilon_\infty (1 + \frac{\nu_{LO}^2 - \nu_{TO}^2}{\nu_{TO}^2 - i\gamma\nu - \nu^2}) \quad (4)$$

where $\nu$ is the frequency in wavenumber, $\varepsilon_\infty = 6.7$ is the high-frequency constant, $\nu_{LO} = 969 \text{ cm}^{-1}$ is the longitudinal optical-phonon frequency, $\nu_{TO} = 793 \text{ cm}^{-1}$ is the transverse optical-phonon frequency, and the scattering rate $\gamma$ equals $4.76 \text{ cm}^{-1}$ at room temperature.

*2.2. Numerical Method*

Spectral-directional emittance of the graphene-covered SiC grating is obtained indirectly as $\varepsilon'_\nu = 1 - R$, where $R$ is the spectral-directional reflectance of the opaque metamaterial structure within the phonon absorption band of SiC. The radiative properties were numerically calculated with the rigorous coupled-wave analysis (RCWA), whose convergence was ensured with a sufficient total of 81 diffraction orders. The thickness of monolayer graphene is considered as $t_G = 0.5$ nm in the calculation, which was verified to be sufficiently small from careful convergence check.

### 3. Results and Discussion

*3.1. Tunable Spectral Normal Emittance with Varying Graphene Chemical Potential*

The spectral emittance at normal direction for transverse-magnetic (TM) polarized wave (i.e., magnetic field is along the grating groove) is plotted in Fig. 2 with varying graphene chemical potential $\mu$. For the bare SiC grating without graphene layer on top, there exists a temporally-coherent emission peak at $\nu_{res} = 853$ cm$^{-1}$ with an peak emittance of 0.73. As studied by Ref. [47], this coherent emission peak is caused by the excitation of phonon-mediated MP inside the SiC grating structure, realized by the collective oscillation of optical phonons or bound charges at the magnetic resonance that form resonant inductor-capacitor (LC) circuitry. The physical mechanism of MP and resulting coherent emission behaviors in the bare SiC gratings have been thoroughly discussed in Ref. [47].

When a graphene sheet with a chemical potential $\mu = 0$ eV is coated onto the SiC grating, the emission peak location barely shifts, but peak emittance increases to 0.96, close to the blackbody emission. When the graphene chemical potential $\mu$ increases from 0 to 1 eV, the emission peak frequency $\nu_{res}$ monotonically shifts from 853 cm$^{-1}$ to 887 cm$^{-1}$, resulting in a relative tunability of 4% in peak frequency. As summarized in Table 1, the quality factor $Q = \nu_{res} / \Delta \nu$ for the emission peaks varies from 31.6 to 42.2 with different $\mu$ values, where $\Delta \nu$ is the peak full width at half maximum.

*3.2. Electromagnetic Field Distribution with and without Monolayer Graphene at MP Resonances*

To explain underlying mechanism responsible for the effect of graphene layer on the coherent emission peak, the electromagnetic (EM) field distributions are plotted for SiC grating structures without and with graphene ($\mu = 0$ eV) respectively in Figs. 3(a) and 3(b) at the same MP resonance frequency $\nu_{res} = 853$

cm$^{-1}$. In the EM field plots, arrows indicate strength and direction for electric field vectors, while contour represents the intensity of magnetic field normalized to the incident wave as $\log_{10}|H/H_o|^2$. Note that the EM field distributions are presented at the cross section of the 1D SiC grating, i.e., the x-z plane.

Figure 3(a) illustrates the EM field distribution for SiC grating without the graphene sheet at resonance frequency $\nu_{res}$ = 853 cm$^{-1}$. It is observed that the electric current oscillates near the surface of SiC around the grating groove, forming a resonant current loop. The magnetic field is significantly enhanced within the groove, with a magnitude of 2 orders stronger than incidence. The EM field pattern presented in Fig. 3(a) distinctly shows the behavior of phonon-mediated MP [47], at which vibration of optical phonons or bound charges in SiC resonates with incident EM field. The resonance induces an oscillating current with significantly enhanced magnetic field inside, and the emission peak arises as a consequence of this diamagnetic response.

Figure 3(b) shows the EM field in the SiC grating structure coated by a monolayer graphene sheet with $\mu$ = 0 eV at same resonance frequency. It can be found that the resonant current loop is also excited, within which the magnetic field is still confined but a little bit weaker inside the grating groove in comparison to that in Fig. 3(a) without graphene monolayer. This is because the graphene sheet is lossy and more optical energy is absorbed by graphene at magnetic resonance. Although a free-standing graphene monolayer has little absorption of 3% or so in the infrared, it could absorb much more when strongly enhanced EM field more than the incidence impinges on the graphene due to the strong localization of electromagnetic energy at magnetic resonance. As absorption is enhanced with graphene-covered SiC grating at the MP resonance, the thermal emission is equivalently strengthened according to the Kirchhoff's law under local thermal equilibrium than the case without graphene. This observation and explanation is also consistent with the study by Zhao et al [48] on the enhanced absorption of a graphene monolayer in the near-infrared due to the magnetic resonance excited inside the cavity of metallic gratings.

*3.3. LC Circuit Model*

As observed in Fig. 3, when MP resonance is excited in the graphene-covered grating, a resonant current is induced at the surfaces around the groove, which can be symbolized by an LC circuit, as shown in Fig. 4(a). The inductance of SiC is determined by $L_{SiC} = L_k + L_m$, where $L_k$ and $L_m$ are respectively the kinetic and mutual parts with expressions as: [47]

$$L_k = -\frac{h'}{\varepsilon_0 \omega^2 \delta} \frac{\varepsilon'_{SiC}}{(\varepsilon'^2_{SiC} + \varepsilon''^2_{SiC})} \quad (5)$$

$$L_m = -\mu_0 h(b+\delta) \quad (6)$$

Note that $\delta = \lambda / 4\pi k$ is the penetration depth of SiC, where $\lambda$ is the wavelength in vacuum and $k$ is the extinction coefficient of SiC. $h'$ is the effective path length that the current flows at the SiC surface. $\varepsilon'_{SiC}$ and $\varepsilon''_{SiC}$ are real and imaginary parts of permittivity of bulk SiC. $\mu_0$ is the vacuum permeability. It should be mentioned that the resonant current is not only oscillating at the very surface of SiC but within a depth of $\delta$. Therefore, we consider that the current oscillates in the central plane with a distance of $\delta/2$ away from SiC surface, which yields $h' = 2h + b + 2\delta$. The vacuum gap in the groove forms a capacitor with capacitance $C_{gap} = c_1 \varepsilon_0 h / b$, where $c_1$ is the coefficient responsible for the non-uniform charge distribution inside the capacitor [47]. Note that, both the effective path length $h'$ and the factor $c_1$ might vary with different geometric parameters and numbers of graphene layers, and thus their expression and values are approximations. $c_1 = 0.5$ is taken as a nominal value considering that the bound charges are linearly distributed at the SiC surfaces and thus treated constant in the present study.

When a graphene layer is attached to the SiC grating, an inductor $L_G$ associated with the graphene sheet should be considered due to the kinetic energy of graphene plasmon. Following the kinetic inductance of SiC in Eq. (5), the inductance of monolayer graphene can be modelled as:

$$L_G = \frac{b+\delta}{\omega} \frac{\sigma''_G}{(\sigma'^2_G + \sigma''^2_G)} \quad (7)$$

where $\sigma'_G$ and $\sigma''_G$ are respectively real and imaginary parts of the graphene conductivity. As shown in Fig. 4(a), the graphene inductor $L_G$ is in parallel with $C_{gap}$. Therefore, the total impedance of the LC circuit becomes:

$$Z_{Total} = Z_{SiC} + Z_G = i\omega L_{SiC} + \frac{i\omega L_G}{1-\omega^2 L_G C_{gap}} \tag{8}$$

The phonon-mediated MP is excited when $Z_{Total} = 0$, which leads to maximum resonance strength. All the inductance, capacitance, and impedance are defined on the per unit length basis along the groove direction. Note that when $\mu = 0$ eV, the graphene has positive real part of permittivity at frequencies larger than $898\,\text{cm}^{-1}$. In this case, the graphene sheet cannot be considered as an inductor but a capacitor instead [49].

The resonance frequency $\nu_{LC}$ predicted by the LC model is calculated for graphene-covered SiC grating structures with $\mu$ varying from 0 to 1 eV. The comparison to the numerical results from the RCWA calculation shows reasonable prediction by the analytical LC model on the MP resonance frequency with a relative difference less than 1.5%. The good agreement on the resonance frequencies of the tunable coherent emission peak between the LC model and RCWA calculation is summarized in Table 1, which undoubtedly confirms the excitation of MP and the dependence of MP frequency on the graphene chemical potential for the novel graphene-covered tunable coherent thermal source.

The tuning effect of graphene chemical potential on the coherent emission frequency associated with MP can be further understood from the LC model. The graphene inductance $L_G$ is strongly dependent on $\mu$, which would ultimately modulate the MP resonance frequency at zero total impedance with $Z_G / Z_{SiC} = -1$ indicated by Eq. (8). To quantitatively explain the increase in resonance frequency with larger graphene chemical potentials, the value of $Z_G / Z_{SiC}$ is plotted in Fig. 4(b) with different $\mu$ values. Note that only $Z_G$ changes with graphene chemical potential, while $Z_{SiC}$ is independent on $\mu$. It is observed that since $Z_G$ and $Z_{SiC}$ are comparable, the change of $Z_G$ induced by varying $\mu$ will greatly shift the MP resonance frequency. It is also found that $Z_G/Z_{SiC}$ decreases with increased $\mu$. Therefore, the resonance frequency of coherent emission peak at which $Z_G/Z_{SiC} = -1$ increases with larger $\mu$ values.

*3.4. Geometrical Dependence of Coherent Emission from Graphene-covered SiC Gratings*

In the light of structural design for practical applications with specific requirement on the coherent emission peak location and strength, the effect of geometric parameters on the coherent emission of the graphene-covered SiC grating is investigated. Figures 5(a), 5(b) and 5(c) are respectively the contour plots of spectral normal emittance as a function of grating height (*h*), groove width (*b*), and grating period (Λ) at TM waves obtained from RCWA calculation. The graphene chemical potential is fixed at $\mu$ = 0.5 eV, and the geometric parameters of the SiC grating are kept at the base values (i.e., Λ = 5 µm, *b* = 0.5 µm, and *h* = 1 µm). As shown in Fig. 5(a), when grating height *h* increases from 0.5 µm to 1.5 µm, the MP resonance peak frequency decreases from $\nu_{res}$ = 910 cm$^{-1}$ to 834 cm$^{-1}$. This is because that, deeper grating grooves with larger *h* values yield increased $C_{gap}$ and $|L_{SiC}|$, which results in increased $Z_G/Z_{SiC}$ values according to the LC model. Different from the effect of grating height, the MP resonance frequency monotonically increases from $\nu_{res}$ = 820 cm$^{-1}$ to 891 cm$^{-1}$ when the groove width *b* increases from 0.1 µm to 1 µm as presented in Fig. 5(b). The effect of groove width *b* on the MP resonance frequency can be explained by the decrease of $Z_G/Z_{SiC}$ values as $C_{gap}$ decreases with *b*. However, the variation of grating period almost does not affect the resonance frequency as shown in Fig. 5(c), simply because grating period has no effect on the MP resonance frequency according to the LC circuit model. The resonance frequencies predicted by LC circuit model for different grating geometries are also plotted as the green triangles, and the good agreement between LC circuit model prediction and RCWA simulation clearly confirms the geometric effects on the MP resonance condition and underlying physical mechanisms. The geometric dependence of the coherent emission from the graphene-covered SiC gratings would also provide guidelines for balancing optimal performance from materials design and manufacturing tolerance in fabrication processes.

*3.5. Angular Dependence of Coherent Emission from Graphene-covered SiC Gratings*

As studied previously, coherent emission due to MP resonance in bare 1D SiC grating structures exhibit directional independence [47]. Therefore, it is worthwhile to investigate the angular behavior and possibly confirm the unique omni-directional thermal emission associated with MP resonance when graphene monolayer is coated onto bare SiC gratings. Figure 6 plots the spectral emittance of graphene-covered SiC gratings as a function of wavenumber $v$ and in-plane wavevector $k_{x0} = (\omega/c_0)\sin\theta$, where $\theta$ is the angle of incidence. The graphene chemical potential is $\mu = 0.5$ eV, and the grating geometry is set as $h = 1\,\mu$m, $b = 0.5\,\mu$m, and $\Lambda = 5$ μm. TM-wave polarization is considered here, only in which the MP could be excited in 1D gratings [47]. A flat selective emission band around $v_{res} = 873$ cm$^{-1}$ is observed in the contour plot, whose physical mechanism is verified as MP resonance by excellent match with the LC model prediction in green triangles (i.e., $v_{LC} = 868$ cm$^{-1}$ for selected angles from 0° to 80°. Therefore, it is confirmed from both numerical simulation and analytical model that, the tunable spectrally-selective thermal emission from the graphene-covered SiC grating also exhibit strong directional independence, which is highly favorable for some applications that require diffuse-like infrared thermal sources.

Besides, there exists a relatively weaker resonance band at higher frequencies, which is associated with the surface modes excited at the vacuum-graphene-SiC grating interface. The dispersion relation of the surface modes can be solved via zeroing the reflection coefficient at the interface given by [33, 45]

$$r^p = \frac{\varepsilon_1 \gamma_0 - \gamma_1 + \sigma_G \gamma_0 \gamma_1 /(\omega \varepsilon_{00})}{\varepsilon_1 \gamma_0 + \gamma_1 + \sigma_G \gamma_0 \gamma_1 /(\omega \varepsilon_{00})} \tag{9}$$

where the subscripts "0" and "1" represent vacuum and SiC medium, respectively. $\sigma_G$ is the graphene conductivity described above. Here, graphene is treated as a sheet current added to the vacuum-SiC interface. $\varepsilon_{00}$ is the absolute dielectric function of vacuum. $\gamma_j = \sqrt{\varepsilon_j \omega^2/c_0^2 - k_{xi}^2}$ is the wavevector component vertical to the interface in medium $j = 0$ or 1. According to the grating function, $k_{xi} = k_{x0} + i2\pi/\Lambda$, where $i$ is the diffraction order. By folding at $k_{x0} = 1/2\Lambda$, i.e., 1000 cm$^{-1}$ for $\Lambda = 5$ μm, the dispersion curve of the surface modes is plotted in Fig. 6, which shows good agreement with the RCWA calcula-tion. Note that the resonance frequencies of the surface modes are highly dependent on $k_{x0}$ or incidence angle $\theta$, which exhibits different behaviors from the direction-independent MP resonance mode.

*3.6. Multilayer Graphene Effect on Tunable Coherent Emission*

In order to possibly achieve a larger tunability on resonance frequency, radiative properties of SiC gratings covered by multiple graphene sheets are further explored. The geometric parameters of SiC gratings are $\Lambda = 5$ μm and $b = 0.5$ μm, while grating height $h$ is changed from 1 μm to 1.5 μm in order to shift the MP resonance frequency at $\mu = 0$ eV to the lower phonon band edge of SiC. In this way, it is attempted to further tune the emission peak to cover most of the phonon absorption band of SiC. The contour plots in Fig. 7 display the calculated spectral normal emittance as a function of $\mu$ from RCWA for the SiC gratings covered with 1, 2, 3 and 4 layers of graphene sheets.

It can be observed that, as the number of graphene layer increases, the tunable spectral range of selective emission peaks increases. Specifically, compared to a monolayer graphene sheet with a tunable range from 820 cm$^{-1}$ to 850 cm$^{-1}$ in Fig. 7(a), double, triple, and quadruple layers of graphene sheets lead to a higher upper limit of the resonance frequencies associated with MP at $\mu = 1$ eV (i.e., 870 cm$^{-1}$, 884 cm$^{-1}$, and 890 cm$^{-1}$, respectively). The lower limit of resonance frequency at $\mu = 0$ eV barely changes with more graphene sheets. As a result, the tunability on the peak emission frequency is improved from 3.7% to 6.1%, 7.8%, and 8.5% when the number of graphene sheets on top of the SiC grating is increased from 1 to 4. Besides, the resonance emission band tends to slightly broaden due to the increased loss with the additional graphene sheets.

The effect of multilayer graphene in further tuning the emission frequency could be understood with the help of the LC model. To account for the effect of multiple graphene sheets coated on top of the SiC gratings, the inductance for multilayer graphene sheets with a total of $m$ layers becomes $L_{G,m} = L_G / m$, where $L_G$ is the inductance of monolayer graphene given by Eq. (7). Here we neglect the inter-coupling between graphene monolayers for simplicity. The impedance of the multiple graphene sheets is then:

$$Z_{G,m} = \frac{i\omega L_G}{m - \omega^2 L_G C_{gap}} \qquad (10)$$

It can be inferred from Eq. (10) that $|Z_{G,m}|$ increases with larger $m$, given the fact that $m - \omega^2 L_G C_{gap}$ is negative in the considered spectral range. When larger $|Z_{G,m}|$ becomes more dominant over $Z_{SiC}$ in the total impedance as $Z_{Total} = Z_{SiC} + Z_{G,m}$, the larger variation of $|Z_{G,m}|$ with multiple graphene sheets will consequently lead to a larger shift of resonance frequency than that with a single layer graphene. The predicted MP resonance frequencies from the LC circuit model at different chemical potentials are presented in Fig. 7 for graphene sheets with different layers. Excellent agreement on the tunable MP resonance frequencies between the RCWA calculations and the analytical LC prediction can be clearly observed. However, when the number of graphene layer further increases, the resonance frequency at large graphene chemical potentials from the RCWA calculation tends to saturate around $v = 900$ cm$^{-1}$, which deviates from the LC prediction at higher resonance frequencies. This is because in graphene-covered SiC grating microstructures, grating-coupled surface modes existing at the air-graphene-SiC interface are mediated by both the graphene plasmon and optical phonons of SiC at high frequencies from $v = 925$ cm$^{-1}$ to 1000 cm$^{-1}$. Therefore, the graphene-tuned MP resonance frequency is suppressed when it approaches the strong surface modes at higher frequencies with larger $\mu$, resulting in altered MP resonance frequencies away from the prediction by the LC model. Note that the simple LC model could not consider the interaction effect between the surface modes and MP resonances.

To further understand the multilayer graphene effect on the MP resonance inside the SiC grating, EM fields are calculated for graphene sheets with different layer numbers at respective MP resonance frequencies, as presented in Fig. 8. The geometric parameters of the SiC grating are the same with those for Fig. 7, while the graphene chemical potential is fixed at $\mu = 0.5$ eV, for which the MP is excited respectively at $v = 835$ cm$^{-1}$, 850 cm$^{-1}$, 862 cm$^{-1}$, or 871 cm$^{-1}$ for 1, 2, 3 or 4 layers of graphene sheets, as indicated in Fig. 7. When graphene layer number increases, it can be observed that the H field strength (shown as contour) around the graphene sheets becomes stronger and decays further into the SiC groove.

This is because at the chemical potential $\mu = 0.5$ eV, graphene acts as an inductor with negative $\varepsilon$ within the phonon absorption band of SiC. As the number of graphene layers increases, larger electrical conductance will result in stronger electrical currents, which leads to stronger H field strength decaying further away from the graphene sheets. On the other hand, the strength of H field confined inside the groove becomes weaker with more graphene layers, due to the fact that less energy can penetrate into the groove with thicker conductive, i.e., more lossy, graphene sheets. Figure 8 clearly illustrates the electromagnetic interaction between multiple graphene layers and the MP excited inside the grooves of SiC gratings. The EM field obtained from the RCWA simulation shows consistency with the LC circuit model depicted in Fig. 4(a) with graphene sheets as an inductor.

## 4. Conclusions

In summary, we have numerically demonstrated a graphene-based spectrally-selective thermal source with tunable emission frequency by modulating graphene chemical potential. The electromagnetic field distribution revealed the MP or magnetic resonance as the physical mechanism that is responsible for the coherent emission. An LC circuit model based on the charge distributions upon magnetic resonance successfully elucidated the mechanism for the modulation effect of graphene chemical potential on the tunable coherent emission. It is shown that the grating height and groove width significantly affect the MP resonance frequency, while the grating period does not. Moreover, the angular independence of the tunable coherent emission from graphene-covered 1D SiC gratings due to MP resonance at TM waves was also demonstrated. By covering the SiC grating with multilayer graphene sheet, the tunable spectral range for the coherent thermal emission can be further broadened to cover most of the phonon absorption band of SiC, demonstrating larger tunabilities on the coherent emission from this novel frequency-tunable infrared thermal source. Insights gained from this work will facilitate the innovative design and wide application of smart IR coherent thermal sources with dynamic spectral tunability.

**Acknowledgments**

This work was supported by a CAREER Award (#1454698) from the National Science Foundation. HW would like to thank the partial support from the US-Australia Solar Energy Collaboration - Micro Urban Solar Integrated Concentrators (MUSIC) project sponsored by the Australian Renewable Energy Agency (ARENA). YY is grateful to the University Graduate Fellowship offered by the ASU Fulton Schools of Engineering. ASU New Faculty Startup fund and Seed Project fund are greatly acknowledged.

**Table 1.** Quality factor for coherent emission peaks with different graphene chemical potential. Resonance frequency obtained from RCWA calculation and LC model are also presented.

| Chemical potential (eV) | 0 | 0.2 | 0.4 | 0.6 | 0.8 | 1 |
|---|---|---|---|---|---|---|
| Quality factor | 31.6 | 39.1 | 36.1 | 39.8 | 40.1 | 42.2 |
| Peak frequency from RCWA, $\nu_{res}$ (cm$^{-1}$) | 853 | 861 | 868 | 875 | 882 | 887 |
| MP frequency from LC model, $\nu_{LC}$ (cm$^{-1}$) | 841 | 850 | 862 | 874 | 886 | 897 |

**Figure Captions:**

**Figure 1.** Schematic of the proposed tunable IR coherent emitter made of graphene-covered SiC gratings.

**Figure 2.** Spectral normal emittance of the tunable coherent emitter at different graphene chemical potential for TM polarized wave. The geometric parameters for the SiC grating are $\Lambda = 5\,\mu m$, $b = 0.5\,\mu m$, and $h = 1\,\mu m$.

**Figure 3.** Electromagnetic fields at MP resonance frequency of $\nu_{res} = 853$ cm$^{-1}$ for SiC grating (a) without graphene and (b) with monolayer graphene at chemical potential $\mu = 0$ eV. The geometric parameters for the SiC grating are $\Lambda = 5\,\mu m$, $b = 0.5\,\mu m$, and $h = 1\,\mu m$.

**Figure 4.** (a) Schematic for the LC circuit model that predicts MP resonance frequency. (b) Value of $Z_G / Z_{SiC}$ in the LC circuit with different graphene chemical potentials.

**Figure 5.** Geometric effects of (a) grating height $h$, (b) groove width $b$, and (c) grating period $\Lambda$ on the spectral normal emittance of graphene-covered SiC gratings at TM waves. The graphene chemical potential is fixed at $\mu = 0.5$ eV. The predicted MP resonance frequencies from the LC circuit model are also plotted as green triangles for comparison with RCWA simulation.

**Figure 6.** Contour plot of the spectral-directional emittance of graphene-covered SiC gratings as a function of wavenumber $\nu$ and in-pane wavevector $k_{x0}$. The parameters are set as $h = 1\,\mu m$, $b = 0.5\,\mu m$, $\Lambda = 5\,\mu m$, and $\mu = 0.5$ eV. The MP resonance frequency is also predicted by the LC circuit model (denoted by green triangles), and the dispersion curve of the surface modes at the vacuum-graphene-SiC interface is also plotted (as blue curves).

**Figure 7.** Spectral normal emittance at TM waves as a function of graphene chemical potential $\mu$ for SiC gratings covered by (a) a single graphene sheet, (b) 2 layers, (c) 3 layers, and (d) 4 layers of graphene sheets. The geometric parameters for the SiC grating are $\Lambda = 5\,\mu m$, $b = 0.5\,\mu m$, and $h = 1.5\,\mu m$.

**Figure 8.** Electromagnetic fields at respective MP resonance frequency for SiC gratings coated by multiple graphene layers with (a) a single graphene sheet at $\nu_{res} = 835$ cm$^{-1}$, (b) 2 layers at $\nu_{res} = 850$ cm$^{-1}$, (c) 3 layers at $\nu_{res} = 862$ cm$^{-1}$, and (d) 4 layers at $\nu_{res} = 871$ cm$^{-1}$. The geometric parameters for the SiC grating are $\Lambda = 5\,\mu m$, $b = 0.5\,\mu m$, and $h = 1.5\,\mu m$. The graphene chemical potential is $\mu = 0.5$ eV.

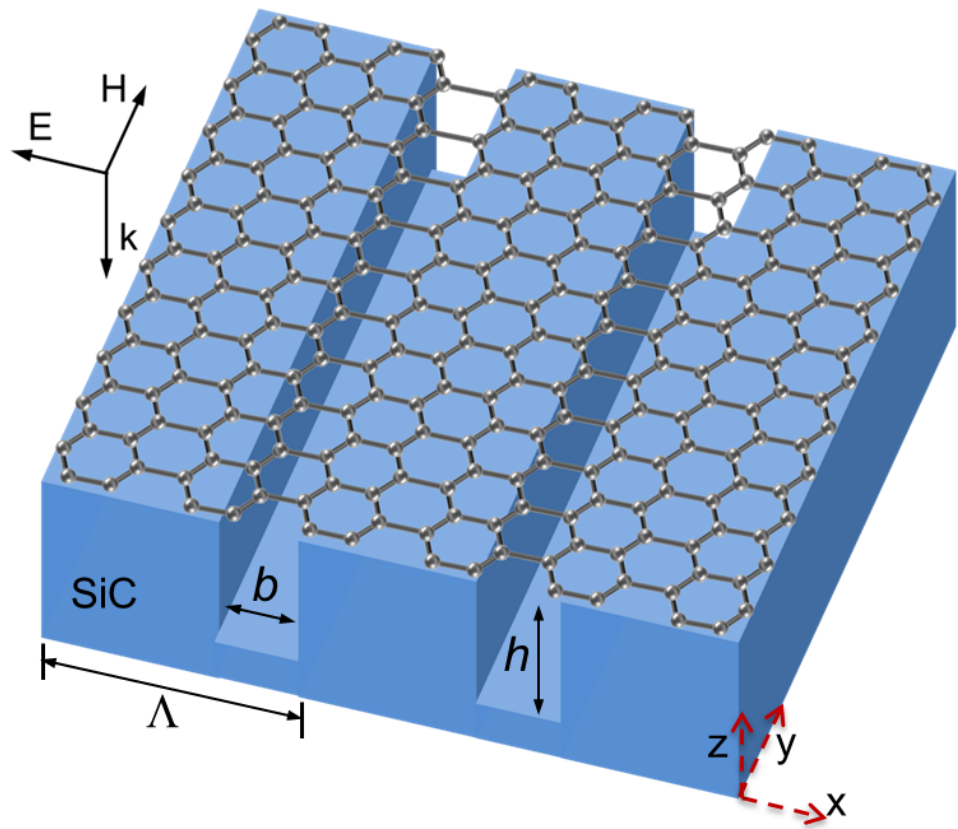

**Figure 1**

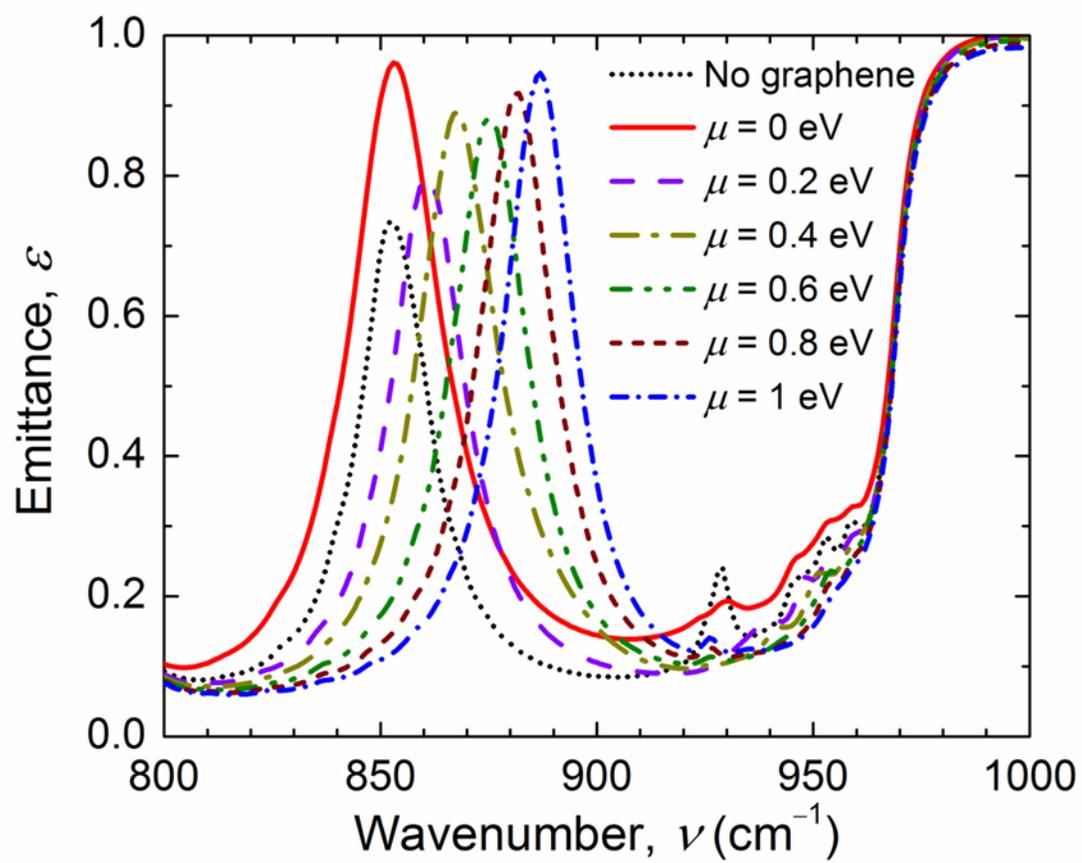

**Figure 2**

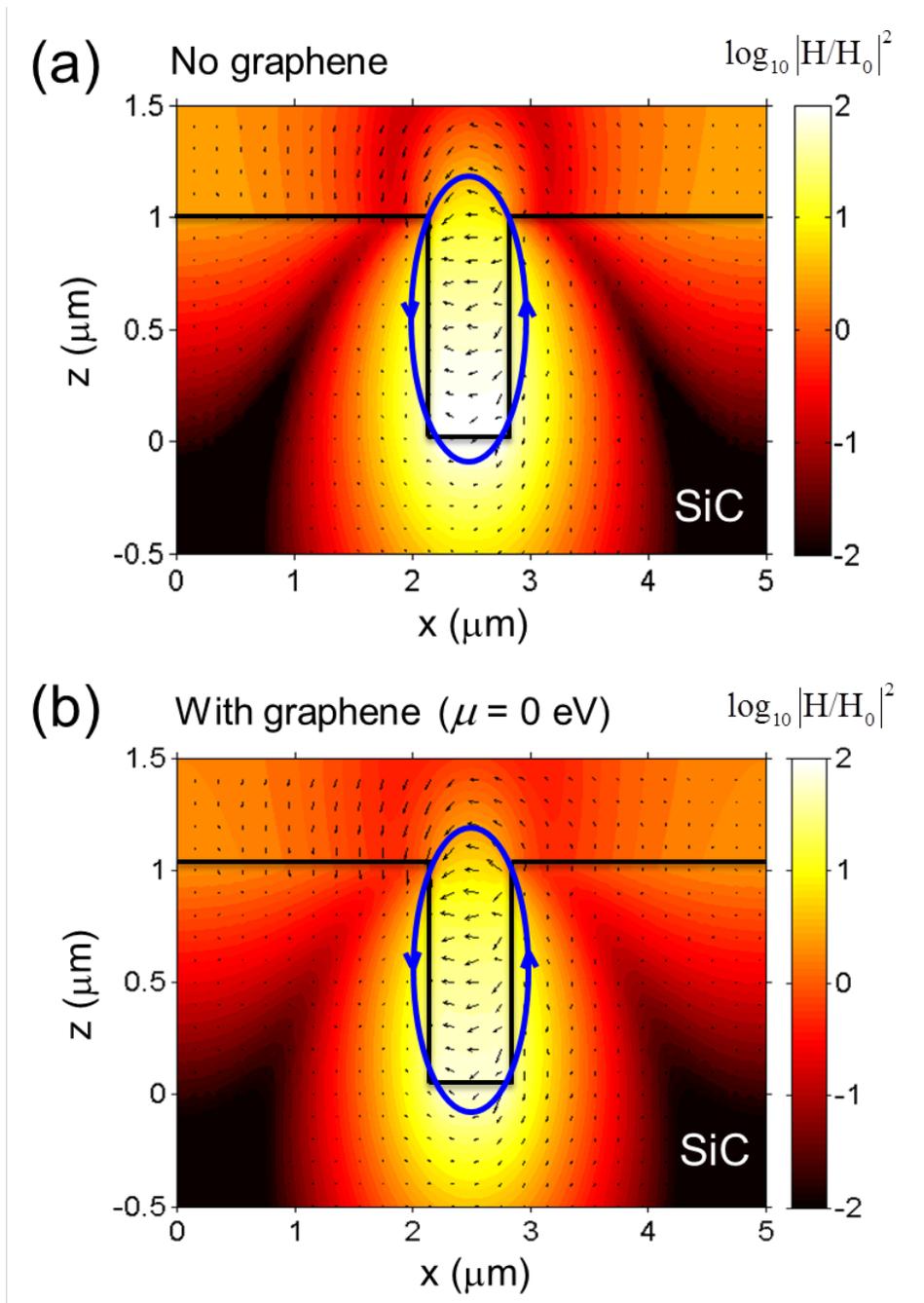

**Figure 3**

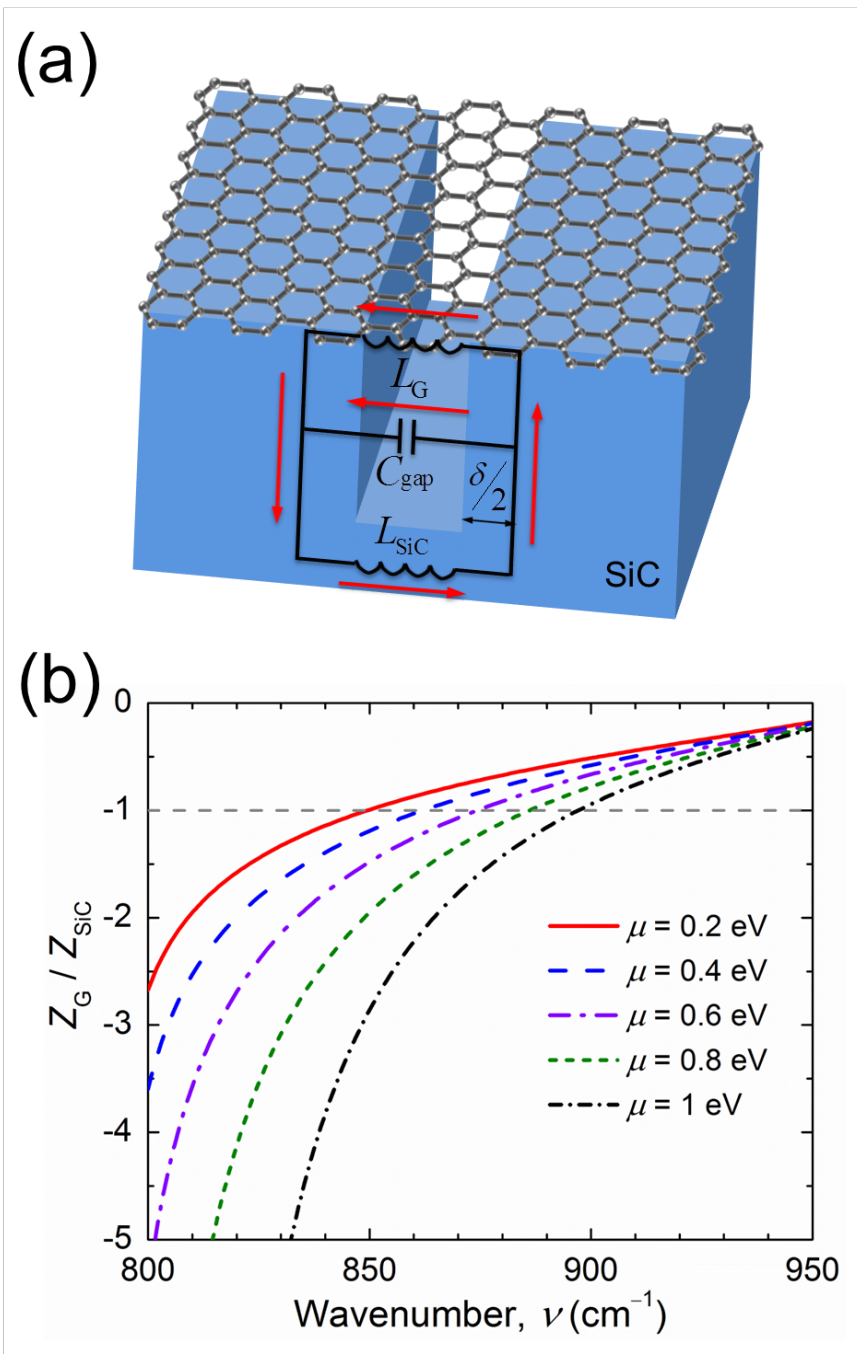

**Figure 4**

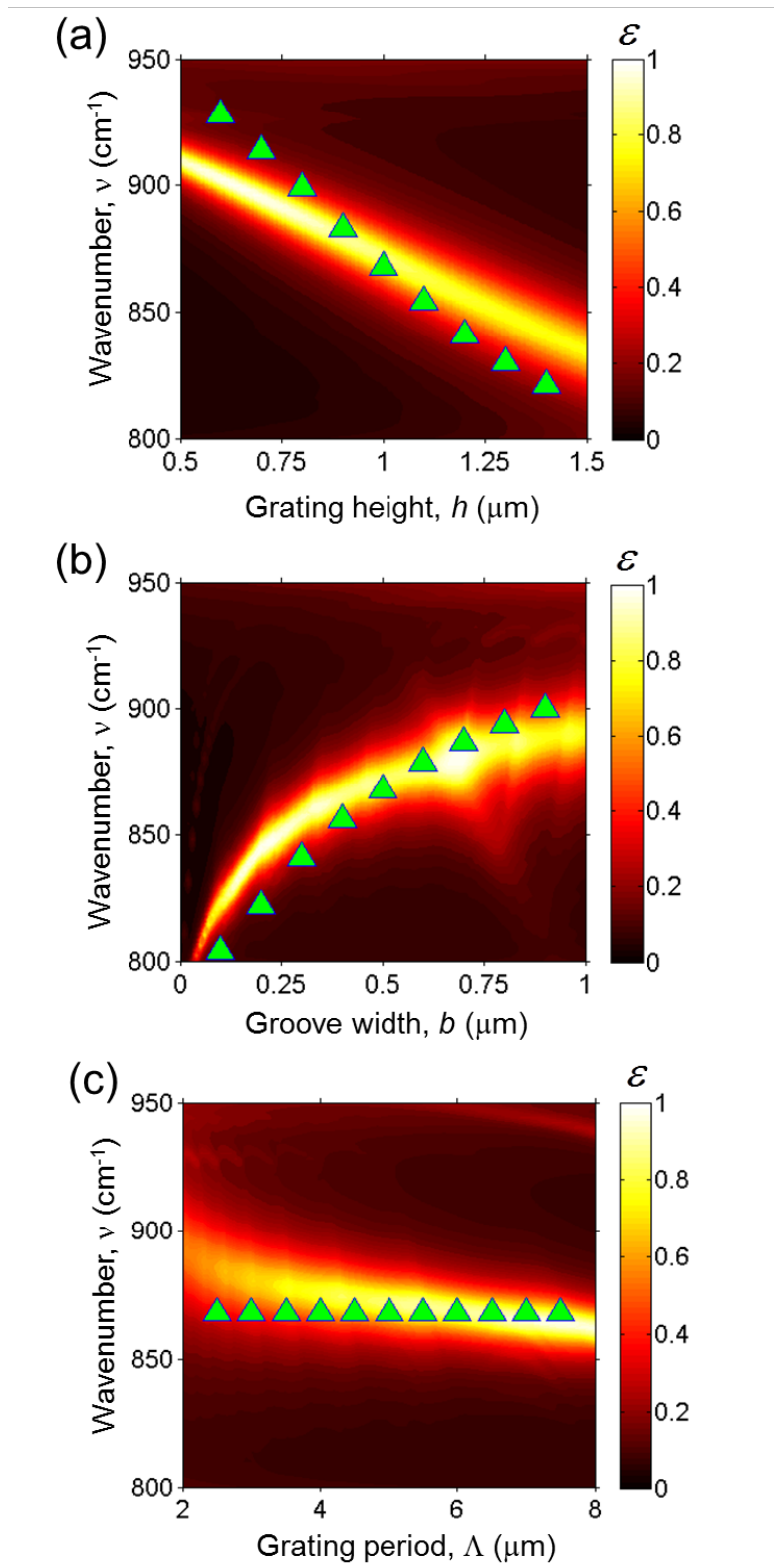

**Figure 5**

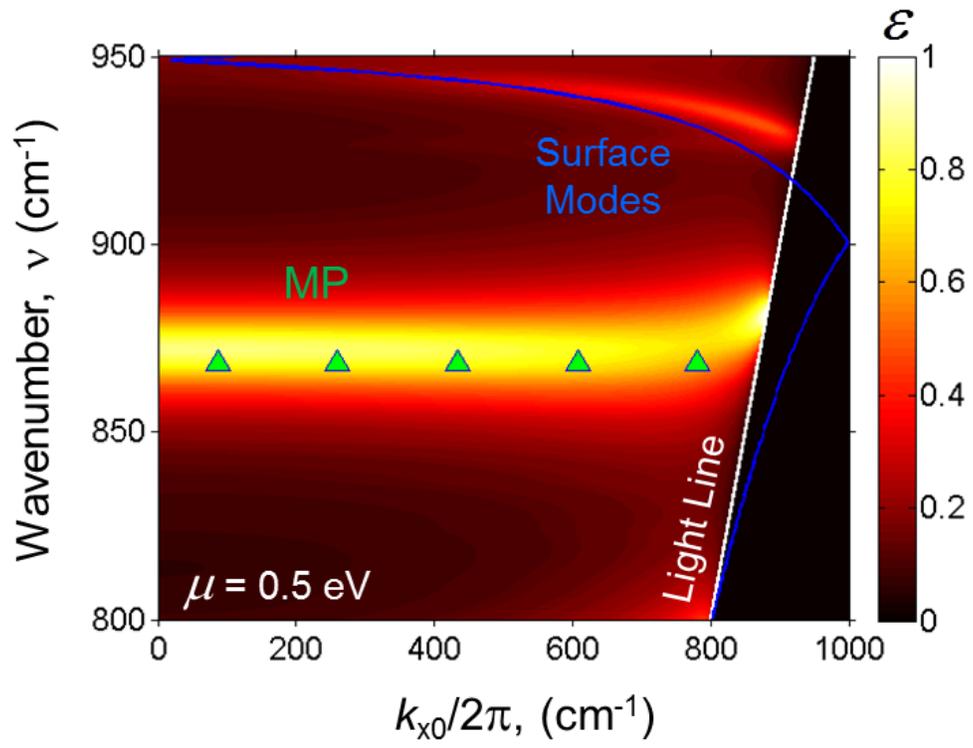

**Figure 6**

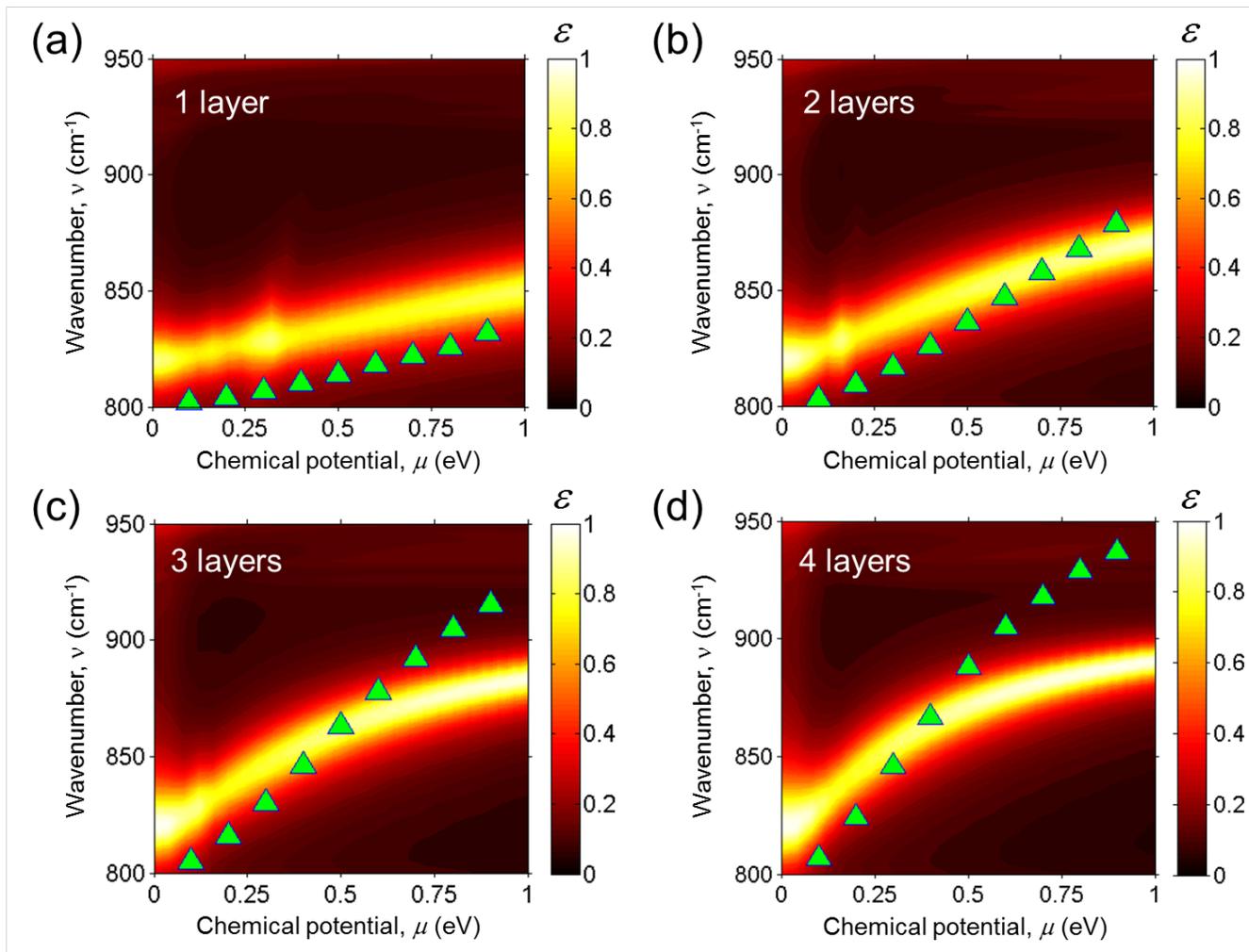

**Figure 7**

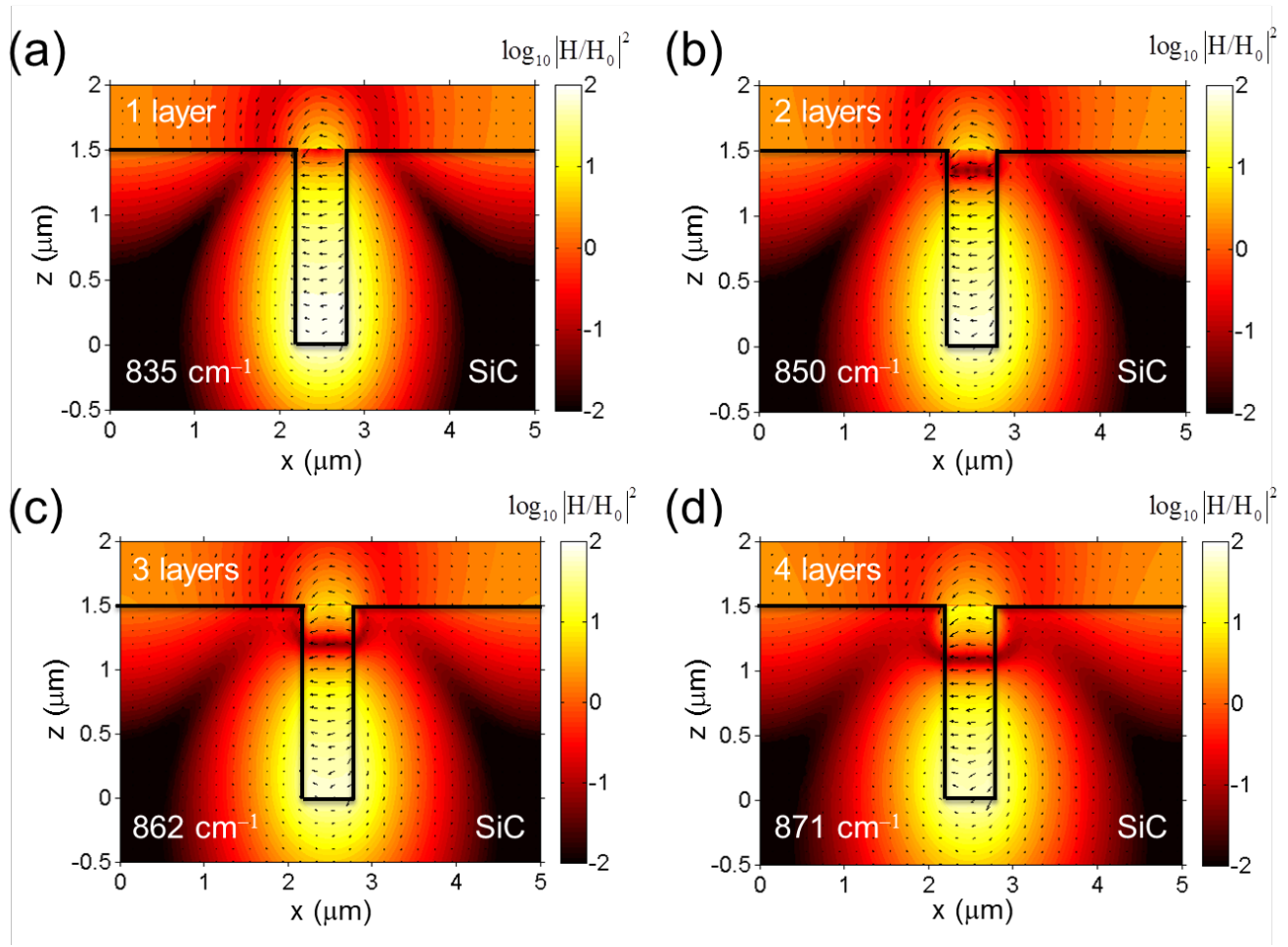

**Figure 8**